\documentclass[11pt,preprint2]{aastex}

\shorttitle{The Shortest Period Non-interacting White Dwarf Binaries}
\shortauthors{Mullally et al.}

\newcommand{\msolar}{\,M$_{\odot}$}

\newcommand{\teff}{$T_{\mathrm{eff}}$}
\newcommand{\logg}{$\log{g}$}

\newcommand{\kms}{\,km.s$^{-1}$}

\newcommand{\wda}{SDSS\,1436}
\newcommand{\wdb}{SDSS\,1053}

\begin{document}
\title{Twins: The Two Shortest Period Non-Interacting Double Degenerate White Dwarf Stars}
\author{F. Mullally\altaffilmark{1}, Carles Badenes\altaffilmark{1}, Susan E. Thompson\altaffilmark{2}, Robert Lupton\altaffilmark{1}}

\altaffiltext{1}{Department of Astrophysical Sciences, Princeton University, Princeton, NJ 08544; fergal@astro.princeton.edu}
\altaffiltext{2}{Department of Physics and Astronomy, University of Delaware, 217 Sharp Lab, Newark, DE 19716, USA}

\begin{abstract}
We report on the detection of the two shortest period non-interacting white dwarf binary systems. These systems, SDSS\,J143633.29+501026.8 and SDSS\,J105353.89+520031.0, were identified by searching for radial velocity variations in the individual exposures that make up the published spectra from the Sloan Digital Sky Survey. We followed up these systems with time series spectroscopy to measure the period and mass ratios of these systems. Although we only place a lower bound on the companion masses, we argue that they must also be white dwarf stars. With periods of approximately 1 hour, we estimate that the systems will merge in less than 100\,Myr, but the merger product will likely not be massive enough to result in a Type 1a supernova.


\end{abstract}


\section{Introduction}
White dwarf stars (WDs) are the end point of stellar evolution for 98\% of all stars \citep{Weidemann00} and store the archaeological record of the Galaxy. WDs in binaries are particularly rich systems to study. Because of their intrinsically low luminosity the companions must also be faint, and are frequently rare or interesting objects. As an example, WDs are ideal targets for the direct detection of planets \citep{Debes05cc2, Farihi08planet, Hogan09, Mullally09} and brown dwarf stars \citep[e.g.][]{Farihi05bd2}. 

The Sloan Digital Sky Survey \citep[SDSS;][]{York00} has increased the number of spectroscopically identified WDs from two thousand to a few tens of thousands. This has, in turn, allowed follow-up surveys of specific types of WD systems, from pulsators \citep[e.g.][]{Mullally05, Nitta09}, to extremely low mass WDs \citep{Kilic07lowmass} to binaries involving main-sequence stars \citep[e.g.][]{Silvestri06, Heller09}, and binaries with neutron stars \citep{Agueros09}.

Binary WDs hold the solution to an enduring problem in astrophysics; the progenitors of Type Ia Supernova (SNIa). The origin of SNIa is of great interest given their role in galactic chemical evolution and determining the nature of dark energy. If a WD accretes enough material that its mass approaches the 
Chandrasekhar limit ($\sim$1.4\msolar), the star can no longer be supported by electron degeneracy pressure and explodes as a supernova. This scenario explains the lack of observed hydrogen in the spectra of SNIa, as well as the striking similarities in the lightcurves and spectra. However, the source of the accreted material, remains a subject of active debate.

In the double degenerate scenario for SNIa progenitors \citep{Iben84, Webbink84}, two WDs in a tight binary in-spiral due to the emission of gravitational radiation. If the total mass is near the Chandrasekhar mass, the merger results in a supernova. While theoretically appealing, it is not clear that nature favors this method.  The SPY survey \citep{Napiwotzki01}, a high precision radial velocity survey of over a thousand WDs, failed to find any candidates with periods short enough to merge within the lifetime of the Galaxy, and masses large enough to explode as SNIa \citep{Nelemans05, Napiwotzki07}.

The SDSS offers an opportunity to build on the SPY survey with a much larger sample of stars. \citet{Kleinman04} and \citet{Eisenstein06} meticulously collected and classified the spectra of nearly 10,000 WD and sub-dwarf stars, of which approximately 8,000 were single stars with hydrogen or helium atmospheres (DAs and DBs respectively). The spectra were obtained as a series of 3 or more 15 minute exposures usually taken consecutively \citep{Abazajian09}, which makes it possible to identify massive companions with orbital periods of a few hours or less and radial velocity amplitudes $\gtrsim$170\kms. The low luminosity of the WD means that any fainter companion must be a degenerate object (WD, brown dwarf, neutron star, etc.) or a very late M star; any other object would be more luminous than the white dwarf.

SWARMS \citep[the Sloan White dwArf Radial velocity data Mining Survey,][]{Badenes09} exploits these individual exposures to mine the SDSS spectroscopic database for double degenerate white dwarf (DDWD) systems. Our survey is complementary to the SPY survey in that it has a lower radial velocity sensitivity, but is still sensitive to white dwarf companions for many thousands of objects. In this paper we present two binary systems, \object{SDSS J143633.29+501026.8} and \object{SDSS J105353.89+520031.0}, with periods of 1.1 and 0.96 hours respectively. These systems constitute the shortest period non-interacting double degenerate binaries yet found, and are significantly shorter than the 1.46 hour period of the previous record holder, \object{WD0957$-$666} \citep{Moran97}. As there are no visible absorption lines from the companions our mass estimates are only lower bounds, but in each case the companion is most likely another WD.

\section{Observations and Reductions}
We identified \wda\ ($g$=18.2, plate-mjd-fiber=1046-52460-594) and \wdb\ ($g$=18.9, 1010-52649-12) as hydrogen atmosphere (DA) WDs potentially possessing short period companions as part of an on-going survey for DDWDs. Although we see radial velocity variations between different exposures, no companion is visible in the spectrum. To confirm these systems as binaries, and to measure the orbital parameters, we observed both stars with the Dual Imaging Spectrograph (DIS) with the 3.5m telescope at Apache Point Observatory over 4 nights between 2009-02-05 and 2009-02-14. We used the B1200 grating with a 1.5" slit for a dispersion of 0.62\,\AA\ per pixel and a resolution of 1.8\,\AA\ FWHM. Each exposure was 10 minutes in duration and bracketed by an exposure of the Helium, Neon and Argon arc lamps. We took several exposures of the spectrophotometric standard Feige~67 each night to flux calibrate our spectra.

\begin{figure}[thb]
    \begin{center}
   \includegraphics[angle=270, scale=.35]{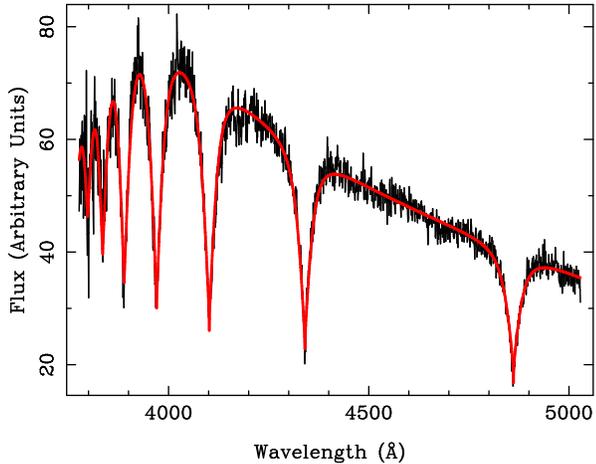}
   \caption{De-shifted and co-added spectrum of \wda\ based on observations made at APO. The solid red line is the best fit model used to estimate the temperature and gravity. The average S/N per pixel in this spectrum is 30, and $\approx$7 in each individual spectra \label{spectrumwda}}
    \end{center}
\end{figure}

We performed an optimal spectroscopic reduction of each spectral image and flux standard using standard long-slit IRAF routines. To optimize the wavelength solution we trace the arc lamp spectra with the same trace used to extract the corresponding WD spectra. A prominent Mercury emission line at 4358\,\AA\ (courteously provided by the residents of White Sands, NM) confirmed our wavelength offset to better than our resolution. The final flux calibrated spectra obtained from the blue {\sc ccd} spans 3790-5020\,\AA. We show an average spectrum from all 4 nights for \wda\ in Figure~\ref{spectrumwda}.

\begin{figure}[thb]
    \begin{center}
   \includegraphics[angle=270, scale=.35]{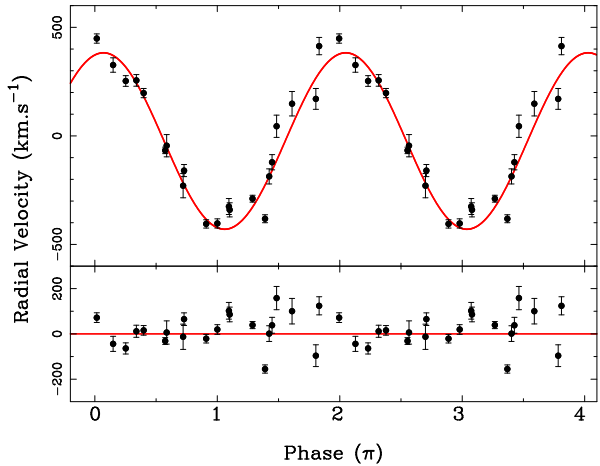}
   \caption{Radial velocity curve for \wda\ folded at the best fit period. The solid line is the best fit sine curve to the data. The residuals of the fit are shown in the lower panel. The apparent linear trend in the residuals is discussed in the text.  \label{periodwda}}
    \end{center}
\end{figure}

\begin{figure}[thb]
    \begin{center}
   \includegraphics[angle=270, scale=.35]{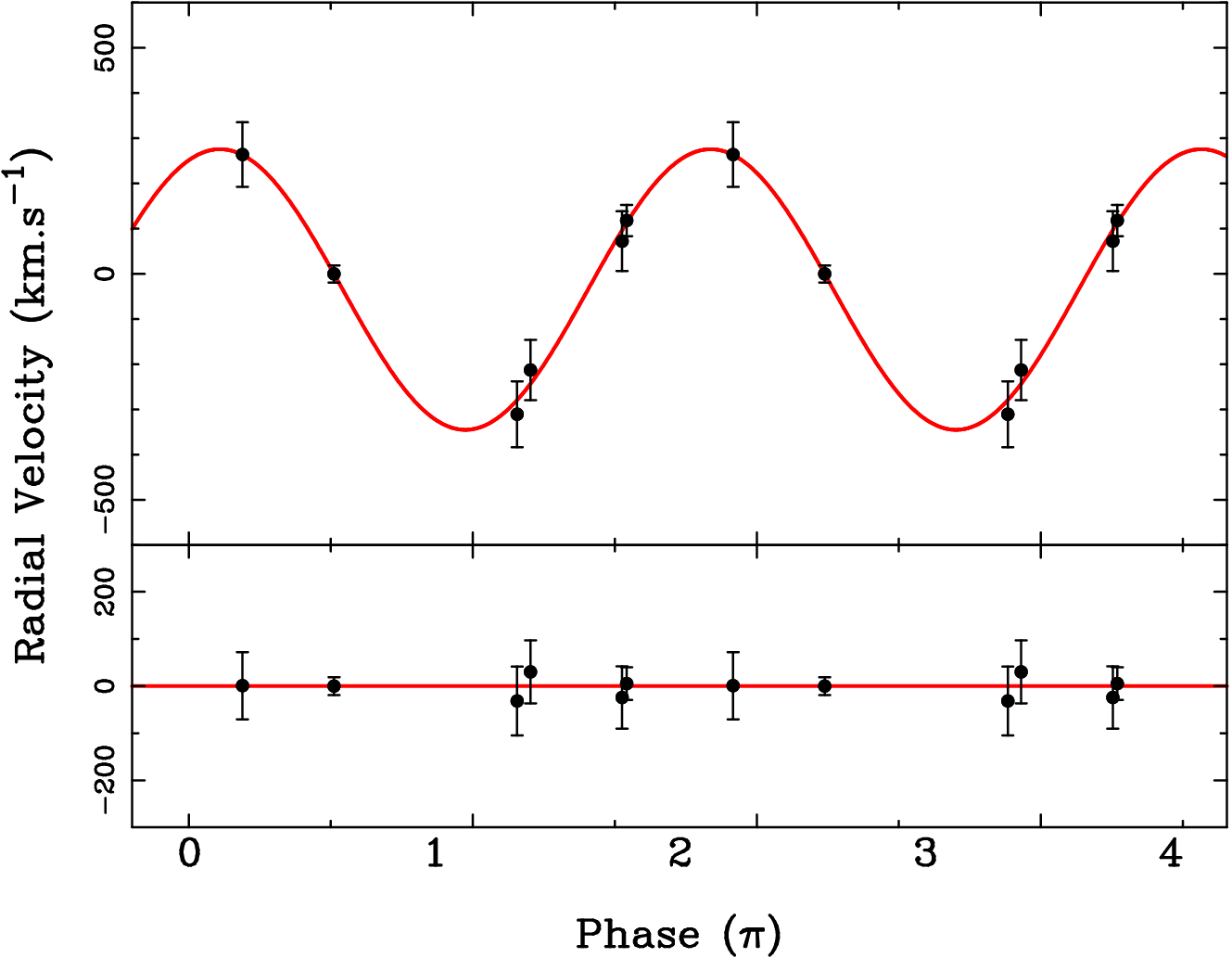}
   \caption{Same as Figure~\ref{periodwda}, but for \wdb. \label{periodwdb}}
    \end{center}
\end{figure}

The pressure broadened lines of a DA white dwarf are well modeled as a combination of a Gaussian core with Lorentz wings \citep{Thompson04}. 
We convert the centroid shift of H$_{\gamma}$ to a velocity, and fit a sine curve with constant offset to the radial velocity time-series. Radial velocity curves for H$_{\beta}$ and H$_{\delta}$ give similar results, but the accuracy obtained by fitting H$_{\gamma}$ alone is sufficient for our purposes. We show the best fit folded radial velocity curves in Figures~\ref{periodwda}~\&~\ref{periodwdb}  and the best fit parameters in Table~\ref{param}. Given the more sparse sampling of \wdb, our period estimate is less certain than for \wda, but our observations span nearly 2 orbits and our uncertainty estimate is only 36s.

The residuals of the fit to \wda\ show a linear trend with phase. This trend is seen for fits to H$_{\beta}$ and H$_{\delta}$ as well. Examination of the unfolded lightcurve confirms this trend is indeed a function of phase, not of time, and can not be explained by some drift in our instrumental calibration. Similarly, a third body in the system on a longer period orbit would only produce a trend in the unfolded data. Fitting an eccentric orbit reduces the peak to peak amplitude of the residual trend by 100\kms\ but does not eliminate it. Because the eccentric orbit fit is not significantly better, and because we have difficulty imagining a scenario in which a system that has undergone two common envelope evolution phases could emerge with an eccentric orbit, we show only the circular fit in Figure~\ref{periodwda}.

\subsection{Temperature and Gravity}
The published temperature of each star from \citet{Eisenstein06} comes from a fit to the average of three separate exposures spanning a total of 50 minutes. As this is a significant fraction of the orbital period we were concerned that the fit may have been biased by combining spectra with different radial velocities. Using our best fit radial velocity curve, we deshifted and co-added each of our 10 minute exposures to produce a high signal-to-noise spectrum. We then fit this spectrum to the same grid of DA models used by \citet{Eisenstein06} \citep[updated by][and kindly provided by the author]{Koester09}. We linearly interpolated the model spectra to produce a finer grid of $\Delta$\teff=10\,K and $\Delta \log{g}$=0.02. We fit each model to the entire spectrum from  3800-5000\,\AA\ using a least squares minimization algorithm, allowing the fit to vary by a high order polynomial in a similar manner to \citet{Eisenstein06}. We find best fit parameters of 17120\,K, \logg=6.60 for \wda\  and 16150\,K, 6.35 for \wdb, consistent with the estimate of \citet{Eisenstein06} who finds (\teff, \logg) of (16933, 6.58) and (15399, 6.28) respectively.

For each object we combine spectra taken close to the minima and maxima of the velocity curve. We find no evidence of spectral features in these combined spectra and are confident that flux from the companion is not biasing our fit.
\citet{Fontaine03} noted that temperature and gravity estimates of WDs from independent spectra using identical reductions and identical atmosphere models often disagree significantly more than the quoted uncertainties. Following their approach, we adopt uncertainties of 200\,K and 0.05 for our fits, which are more conservative than the values returned by the fitting method. We caution that these uncertainties are internal to our fitting, and do not attempt to address limitations of the models. For example, \citet{Tremblay09} recently introduced an improved treatment of Stark broadening which systematically increases the best fit gravity by 0.2 dex in this temperature and gravity range.

\citet{Kilic07lowmass} independently observed and fit the spectra of both stars (as part of a search for companions to low mass WDs) and obtained similar results for the gravity (\logg = 6.59 and 6.40), but higher temperatures (\teff = 18339 and 18325). Given the close agreement in measured gravity between the three measurements, the small discrepancy in temperatures do not materially effect the stellar masses we estimate in Section~\ref{discussion}




We simulated the effect of changing radial velocities over the course of a 10 minute exposure to estimate the effect on the best fit temperature and gravity. Using our best fit radial velocity curve for \wda, we coadded a series of appropriately Doppler shifted model spectra, and fit the result in a manner similar to our data. The largest discrepancy occurs when the star is traveling perpendicular to the line of sight, where the core of the line appears smoothed. This blurred spectrum is preferentially fit by 500\,K hotter model with a shallower line core, but the best fit gravity, which is most important for measuring the mass, remains unchanged.

\section{Discussion \label{discussion}}
Comparing our best fit temperature and gravity to the WD evolution models of \citet{Serenelli02} we estimate masses of 0.23(01) and 0.21(01)\msolar. These models are created by removing mass from a 1\msolar\ model at appropriate times during red giant branch evolution, and incorporate chemical diffusion, a nearly pure He core (with metallicity, Z=0.001) and thick H layer. The estimated masses are close to the minimum known white dwarf mass of 0.17\msolar\ \citep{Kilic07lowmass, Kawka09}. 
\citet{Moroni09} estimates that a WD must have a mass of at least 0.33\msolar\ to have a carbon-oxygen core. Neither object approaches this mass, and are composed almost entirely of helium with a thin hydrogen atmosphere.

\begin{figure}[!thb]
    \begin{center}
   \includegraphics[angle=90, scale=.85]{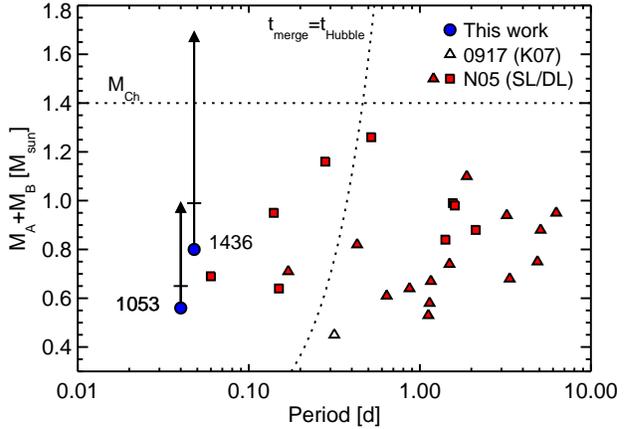}
   \caption{
Distribution of known DDWD systems. The filled circles indicate the minimum mass of the two new systems discussed in this paper. The notch on the arrow gives the total mass assuming an inclination angle, $\iota$, of 60$^{\circ}$, and the tip of the arrow shows $\iota=40^{\circ}$. Square symbols indicate previously known double-lined (DL) binaries (where the total mass is known), while triangles indicate single-lined (SL) systems and are only a lower bound on the total system mass. 
\protect\object{SDSS\,J091709.55+463821.8} (open triangle) taken from 
\citet{Kilic07lowmasscomp}, and all other systems from \citet{Nelemans05}. The horizontal dashed line indicates the Chandrasekhar mass, while the curved line shows the period for which the merger time is equal to the age of the Universe.
   \label{mergetime}
    }
    \end{center}
\end{figure}

According to the initial-final mass relation \citep[e.g.][]{Kalirai07, Williams09}, only isolated WDs with masses greater than 0.47\msolar\ have had time to evolve off the main-sequence within the lifetime of the Universe. Although it has been argued that high metallicity progenitors can produce lower mass WDs \citep{Kilic07future}, that scenario is unnecessary for these systems. Instead, the fact that these two systems are known to be binaries suggests that growth of these lower mass WDs was truncated by a common envelope phase and evolved through the sub-dwarf channel \citep{Heber09}.

\begin{deluxetable}{lrr}
\tablewidth{0pt}
\tablecaption{Measured Parameters \label{param}}

\tablehead{
    \colhead{}&
    \colhead{\wda}&
    \colhead{\wdb}
}

\startdata
Period (hrs)                  & 1.15238(14) & 0.960(10) \\
Amplitude (km.s$^{-1}$)       & 388(21)     & 310(14) \\
Seperation  ($R_{\odot}$)     & 0.4789(75)  & 0.2924(51) \\
\teff (K)                     & 17120(200)  & 16150(200)    \\                 
\logg                         & 6.60(05)   & 6.35(05) \\
Mass$_1$ (M$_{\odot}$)        & 0.23(01)     & 0.22(01) \\
Mass$_2/\sin{i}$ (M$_{\odot}$) & 0.57(04)          & 0.31(02) \\
Merge time (Myr)            & $<$102          & $<$104 \\

\enddata 
 
\end{deluxetable}

\subsection{Nature of the Companions}
%

Solving for the Keplerian equations of motion for \wda\ gives a mass for the companion of (0.57(04)/$\sin{\iota}$)\msolar, where $\iota$ is the inclination of the orbit to the line of sight, consistent with a carbon-oxygen core WD. The companion to \wdb\ is at least 0.31(02)\msolar. 
Although these are minimum masses, and are consistent with a wide range of astrophysical objects, we argue that the companions are most likely also WDs.

Main sequence stars can be ruled out on luminosity grounds. For \wda\ (\wdb), the minimum mass of the companion corresponds to a spectral type of K8 (M1) \citep{Habets81}, which has an absolute $i$ magnitude of 7.2 (8.5) \citep{Bilir09,Hawley02}, considerably brighter than the observed WD \citep[$i = 9.1~(9.4)$,][]{Holberg06}. The SDSS spectrum of either object shows no evidence of any cool companion at red wavelengths, ruling out the possibility of a main-sequence companion. A similar argument applies to red giant stars and other, higher luminosity objects.

If the inclination angle, $\iota < 24^{\circ} (13^{\circ})$, the companion mass is greater than the Chandrasekhar mass and the companion must be a neutron star (NS) or a black hole. Approximately 45 WD-NS binaries are known, and the mass distribution of WDs in such binaries is much wider than for isolated systems, admitting both high and low mass WDs \citep{vanKerkwijk05}. 
\citet{Agueros09} looked at both \wda\ and \wdb\ during an 820\,Mhz radio survey for pulsar companions to WDs but did not detect any signal. These observations do not exclude the possibility of a pulsar companion, not only because the orientation of the pulsar beam may not be along the line of sight, but also because their analysis restricted their sensitivity to orbital periods greater than 8 hours.

Interacting WD-NS binaries are known as ultra compact X-ray binaries \citep[UCXBs; see][for a review]{Nelemans06}. In these systems, the orbital separation is so small that a WD overfills its Roche lobe and donates material onto the surface of the neutron star via an accretion disk, emitting X-rays in the process. The longest period UXCBs have periods of 50-55 minutes \citep{Nelemans06}, entirely consistent with the periods of the systems under scrutiny. If the companions were NSs, they would almost certainly be interacting.
However, the presence of hydrogen in the atmosphere of the visible WD in both systems  means that the systems are {\it not} interacting. In double degenerate systems, mass transfers from the lower to the higher mass star. In both systems, the higher mass star is the invisible companion. If the visible star was losing mass, the thin hydrogen layer would be quickly stripped, exposing the underlying helium core. Because both stars still have their hydrogen layers we can conclude that mass transfer has not yet started.

\subsection{Consequences of a Merger}
Non-interacting DDWD companions are therefore the only possible objects consistent with the available evidence. Non-interacting systems in short period orbits lose orbital energy in the form of gravitational radiation and will eventually merge.  Using the equation for angular momentum loss given by \citet{Paczynski67}, we estimate these systems will merge in less than 102 and 104\,Myr respectively. If the orbits are inclined by 45$^{\circ}$ to the line of sight, the merger time decreases to $<79$\,Myr

\citet{Guerrero04} used smoothed particle hydrodynamic simulations to predict the consequences of the merger of WDs of different masses. For the merger of two 0.4\msolar\ helium core WDs (analogous to the minimum mass configuration of the 
\wdb\ system), they found no thermonuclear flash, and no mass loss. The case of a 0.4\msolar\ WD merging with a 0.6\msolar\ carbon-oxygen core star (similar to \wda), some carbon is burned into oxygen, but there is no thermonuclear runaway and no supernova.

Unless the unseen companions have masses $\gtrsim 1.2$\msolar\ it seems unlikely that either system will produce a SNIa. WDs with masses greater than 1.2\msolar\ do exist, with the most massive WD found to date being 1.33\msolar\ \citep[][assuming an oxygen-neon core]{Kepler07}. However, these stars are rare, and are consistent with only a small range of inclinations angles for these systems. 

The merger of a carbon-oxygen core WD with a helium WD most likely produces an extreme helium star \citep{Saio02} or an R CrB star \citep{Webbink84, Clayton07}. The merger of two helium core WDs (an option only for \wdb) is one evolutionary pathway to produce sdO subdwarf stars \citep{Heber09}. 
Regardless, the merger remnant will eventually cool to become a single WD. \citet{Liebert05} compared the space density of high mass WDs with that expected from a Salpeter initial mass function and a single burst stellar population and concluded that 80\% of high mass WDs were created by the merger of lower mass stars. The merger of a carbon-oxygen WD with a He core one has been suggested by \citet{Garcia07} as the origin of hot debris disks around massive WDs. 

However, analysis of the extremely low luminosity, and calcium rich, type 1b supernova SN2008E by \citet{Perets09} concluded that only 0.3\msolar\ of material was ejected, and the pattern of elemental abundances (high calcium abundances, but low sulpher) was best explained by helium fraction $>$0.5 in the initial composition. It is conceivable that the progenitor of this explosion involved the disruption of a helium core WD in a DDWD system.

\section{Conclusion}
We report on the detection of the two closest non-contact  white dwarf binaries known. These systems were detected by mining the spectroscopic database of the SDSS, and followed-up with time resolved optical spectroscopy.  We argue that the companions must also be WDs:
Main-sequence stars of the requisite mass are more luminous than the primary WDs, and a neutron star or black hole in such close proximity would have stripped off the thin outer layers of the primaries. With periods of about an hour, these systems will merge in less than 100\,Myr and probably produce a high mass WD.

\acknowledgements
 CB is supported by NASA through Chandra Postdoctoral Fellowship Award Number PF6-70046 issued by
  the Chandra X-ray Observatory Center, which is operated by the Smithsonian Astrophysical Observatory for and on behalf
  of NASA under contract NAS8-03060. SET acknowledges the support of the Crystal Trust. We wish to thank the staff of Apache Point Observatory for their assistance in making these observations.


\onecolumn


\end{document}